\documentclass[aoas,nameyear]{arximspdf}
\usepackage{dcolumn}
\usepackage{graphicx}

\doi{10.1214/09-AOAS291}
\volume{3}
\issue{4}
\pubyear{2009}
\firstpage{1309}
\lastpage{1334}

\makeatletter
\newcolumntype{d}[1]{D{.}{.}{#1}}

\def\textttm#1{\texttt{#1}}
\makeatother

\begin{document}
\begin{frontmatter}

\title{Deriving chemosensitivity from cell lines:\\
Forensic bioinformatics and reproducible\\
research in high-throughput biology}
\runtitle{Cell Lines, Chemo and Reproducibility}

\begin{aug}
\author{\fnms{Keith A.} \snm{Baggerly}\corref{}\thanksref{t1}\ead[label=e1]{kabagg@mdanderson.org}}\and
\author{\fnms{Kevin R.} \snm{Coombes}\thanksref{t2}\ead[label=e2]{krcoombes@mdanderson.org}}
\thankstext{t1}{Supported in part by NIH Grant P50 CA083639 09.}
\thankstext{t2}{Supported in part by NIH Grant P50 CA070907 08.}

\runauthor{K.~A.~Baggerly and K.~R.~Coombes}
\affiliation{University of Texas}
\address{Department of Bioinformatics\\
\quad and Computational Biology\\
M. D. Anderson Cancer Center\\
University of Texas\\
1400 Pressler\\
Houston, Texas 77030\\
USA\\
\printead{e1}\\
\phantom{E-mail: }\printead*{e2}} 
\end{aug}

\received{\smonth{9} \syear{2009}}

\begin{abstract}
High-throughput biological assays such as microarrays
let us ask very detailed questions about how diseases
operate, and promise to let us personalize therapy.
Data processing, however, is often not described well
enough to allow for exact reproduction of the results,
leading to exercises in ``forensic bioinformatics''
where aspects of raw data and reported results are
used to infer what methods must have been employed.
Unfortunately, poor documentation can shift from an inconvenience
to an active danger when it obscures not just methods but errors.
In this report we examine several related papers purporting
to use microarray-based signatures of drug sensitivity
derived from cell lines to predict patient response.
Patients in clinical trials are currently being allocated
to treatment arms on the basis of these results. However,
we show in five case studies that the results incorporate
several simple errors that may be putting patients at risk.
One theme that emerges is that the most common errors are
simple (e.g., row or column offsets); conversely, it is
our experience that the most simple errors are common.
We then discuss steps we are taking to avoid such errors
in our own investigations.
\end{abstract}


\end{frontmatter}

\section{Background}\label{sec1}

\subsection{General}

High-throughput biological assays such as microarrays
let us ask very detailed questions about how diseases
operate, and promise to let us personalize therapy.
For example, if we know a priori that only about
70\% of a group of cancer patients will respond to
standard front-line therapy (as is the case with ovarian
cancer), then an array-based test indicating whether
a given patient is likely to respond would have great
clinical utility.

Data processing in high-throughput studies, however,
is often not described well enough to allow for exact
reproduction of the results. While various groups have
made efforts to construct tools or compendia that
should make such reproducibility easier
[e.g.,
\citet{leisch02},
\citet{ruschhaupt04},
\citet{gentleman05},
\citet{gentleman07},
\citet{li08}],
a recent survey [\citet{ioannidis09}]
of 18 quantitative papers
published in \textit{Nature Genetics} in the past two years
found 
reproducibility was not achievable even in principle for 10.
The time period is relevant, as the journal
was then requiring that all raw data be deposited in
public repositories so
reproducibility
would be easier.
In some cases, more extensive exercises in ``forensic bioinformatics''
[e.g.,
\citet{stivers03},
\citet{baggerly04a},
\citeauthor{baggerly04b} (\citeyear{baggerly04b,baggerly05}),
\citet{coombes07}]
can use aspects of raw data and reported results
to infer the methods that must have been employed,
but these are 
time-consuming.

Unfortunately, poor documentation and irreproducibility
can shift from an inconvenience to an active danger when
it obscures not just methods but errors. This can lead to
scenarios where well-meaning investigators argue in good
faith for treating patients with apparently promising
drugs that are in fact ineffective or even contraindicated.
In some ways, this problem is qualitatively worse for
high-throughput assays than it is for simple tests,
because the sheer magnitude of the data confounds both
our ability to spot simple errors and our intuition about
how certain tests ``should'' behave. For several single-marker
assays, investigators may have a well-developed intuition
that high values indicate a poor prognosis, but 50-gene
signatures require faith that the assembly code has been
applied correctly.

In this report we try to illustrate the potential severity
of the problem by examining several related papers purporting
to use microarray-based signatures of drug sensitivity
derived from cell lines to predict patient response.
Patients in clinical trials are currently being allocated
to treatment arms on the basis of these results. However,
we show in five case studies that the results incorporate
several simple errors that may be putting patients at risk.
The complete code and documentation underlying our results
are available as supplementary material (see the Appendix).
We then discuss steps we are taking to avoid such errors
in our own investigations.

\subsection{Chemosensitivity and cell lines}

While many microarray studies have clarified aspects of
basic cancer biology, there is a definite push to make
our studies more ``translational'' in that our bench or
in silico results should be likely to translate into
changes in patient care in short order. To that end,
studies identifying new targets for drug development
are of less import than studies guiding how we use the
drugs available to us now. In short, we would like to
know whether a given patient is likely to be sensitive
(aka~be a responder) to a given agent.

In this context, we briefly consider how the activity of
new chemotherapeutics is assessed. Shortly after
cytotoxic chemotherapy was introduced for the treatment
of leukemia, the U.S. government established the
Cancer Chemotherapy National Service Center (CCNSC)
in part to help test new agents. Drugs were submitted,
assigned an NSC number, and 
used to treat leukemic
mice to see if they produced improvement. Both the agency
and the method of testing have changed over time.
The CCNSC was absorbed into the Developmental Therapeutics
Program (DTP), but every new drug submitted is still assigned
an NSC number. Tests shifted first from leukemic mice to
immunologically compromised (nude) mice onto which various
types of human tumors were grafted, and thence to a standard
panel of cell lines (the NCI60) derived from 9 
types of human tumors. Sensitivity is assessed in cell
lines through dilution assays. Six samples
(aliquots) of a given cell line are put into separate wells,
a starting dose is applied to well 1, one tenth that dose
is applied to well 2, and so on through well 5; well 6 is left untreated.
After a fixed 
period, the amount of growth seen in
the treated wells is assessed relative to 
growth in the
untreated well. Summary measures reported are the estimated
doses required to achieve 50\% growth inhibition (GI50 or IC50),
total growth inhibition (TGI) or not only growth inhibition
but 50\% lethality of the starting cells (LC50). The important
thing about this approach is that data on activity against the
NCI60 panel is publicly available for every chemotherapeutic
now in use, so a method of constructing sensitivity signatures
from this panel would have extremely broad applicability.

\subsection{Initial claims}
In late 2006, \citet{potti06}
introduced a
method for combining microarray profiles of the NCI60
with drug sensitivity data to derive ``signatures'' of
sensitivity to specific drugs, which they then
used to predict patient response. In theory, the approach is
straightforward:
\begin{itemize}
  \item Using drug sensitivity data for a panel of
    cell lines, choose those that are most sensitive
    and most resistant to a drug.
  \item Using array profiles of the chosen
    cell lines, select the most differentially
    expressed genes.
  \item Using the selected genes,
    build a model that takes an array profile
    and returns a classification, and use this
    model to predict patient response.
\end{itemize}
They reported success using this approach with several
common chemotherapeutic agents, specifically docetaxel,
doxorubicin (adriamycin), paclitaxel (taxol), 5-fluorouracil
(fluorouracil, FU, or 5-FU), cyclophosphamide (cytoxan),
etoposide and topotecan. They also reported some initial
success at predicting response to combination therapies.
Unsurprisingly, these results generated a lot of attention.
The approach was named
one of ``The Top 6 Genetics Stories
of 2006'' [\citet{discover07}],
is the
subject of U.S. Patent Application 20090105167, ``Predicting
responsiveness to cancer therapeutics''
[\citet{uspatent09}],
and had been cited
in 212
papers by August 2009 (Google Scholar).

\subsection{Initial questions and response}
Several groups at MD Anderson were among those excited,
and we examined the approach in order to help our
investigators use this.
However, as documented in \citet{coombes07},
when we
independently reanalyzed the raw data, we noted a number of
simple errors affecting the conclusions.
Taking doxorubicin as an example, we noted
that \citet{potti06}
predicted response
in a test set cohort involving samples from 122 patients;
23 patients were reportedly sensitive
to doxorubicin and 99 resistant [\citet{potti06},
Figure~2C]. However, this data was said to arise from an
``independent dataset of samples cultured from adriamycin-treated
individuals (GEO accession nos.~GSE650 and GSE651).''
These Gene Expression Omnibus (GEO) data sets, which
derive from a study on acute lymphocytic
leukemia (ALL) by \citet{holleman04},
give gene profiles for samples from 94 patients sensitive to
daunorubicin (in the same chemical family as doxorubicin) and 28 resistant.
While the total number of patients (122) is constant, the
proportions ($23{}\dvtx{}99$ and $94{}\dvtx{}28$) are almost inverted, which led
us to suggest that the
sensitive/resistant labels might have been reversed.
Label reversal implies that an ``accurate'' signature
would preferentially suggest doxorubicin for patients
who would \textit{not} benefit.

In reply,
\citet{potti07}
disagree,
claiming that their approach is ``reproducible and robust.''
In support of this claim, they comment on ``the acute lymphocytic
leukemia dataset in which the labels are accurate---full details
are provided on our web page.'' They note having gotten the
approach to work again more recently
[e.g., \citet{hsu07}, \citet{bonnefoi07}],
and conclude that the \citet{coombes07}
analysis is flawed.

\subsection{Subsequent progress}
Progress since 2006 appears substantial.\break
\citet{hsu07}
used the same approach
to develop signatures of response to cisplatin and pemetrexed,
extending the range of treatments that could be compared.
In examining these signatures, they identified specific
components strongly suggesting biological plausibility, noting that

\begin{quote}
``The cisplatin sensitivity predictor
includes DNA repair genes such as ERCC1 and ERCC4, among others,
that had altered expression in the list of cisplatin sensitivity
predictor genes. Interestingly, one previously described mechanism
of resistance to cisplatin therapy results from the increased
capacity of cancer cells to repair DNA damage incurred, by
activation of DNA repair genes.''
\end{quote}

\noindent Clinical trial TOP0602 (NCT00509366), now underway, will
``assign subjects to either pemetrexed/gemcitabine or
cisplatin/gemcitabine therapy using a genomic based platinum
predictor to determine chemotherapy sensitivity and predict
response to chemotherapy for first-line therapy in
advanced non-small cell lung cancer.''
[\citet{clintrial09}].

Later,
\citet{bonnefoi07}
provided a ``validation''
of the combination approach, using it to predict patient response
to two alternative therapies: taxotere followed by epirubicin
and taxotere (TET), and fluorouracil, epirubicin and cyclophosphamide
(FEC). This report is a substudy of the European
clinical trial EORTC 10994/BIG 00-01, in which breast cancer patients
are randomized to receive either TET or FEC. 
Based on their results,
\citet{bonnefoi07}
suggest that using their predictions might have
improved response rates in the target population from 44\% to 70\%,
a huge benefit.

The most recent application was
when
\citet{augustine09}
used the
approach to construct a signature for temozolomide.
The procedure now appears fairly standardized:

\begin{quote}
``A signature of gene expression that correlated with resistance to
temozolomide was derived from the NCI-60 panel of cancer
cell lines (ref.~17; see also Supplementary data). From this
panel of 60 cell lines, a smaller subset of 15 was selected that
represented two extremes of sensitivity to temozolomide; nine
of these cell lines were classified as resistant and six as
sensitive. Using the gene expression profiles of these cell
lines, we identified 45 genes that showed significantly different
expression patterns between the resistant and sensitive cell
lines and thus provided a temozolomide sensitivity gene
signature (see Supplementary Table S4). The color-coded
heatmap of expression of the 45 temozolomide genes across
these 15 cell lines (Fig. 4A) shows 8 genes (red) that were more
highly expressed in the resistant than in the sensitive cell lines,
whereas 37 genes (blue) were more highly expressed in the
sensitive than in the resistant cell lines.''
\end{quote}

In sum, the procedure apparently gives good predictions
in independent test sets, has some biological plausibility,
appears to be giving stable results over years of application,
and is consequently guiding treatment.

\subsection{Cases we examine here}
While there are other studies, we examine four of the cases
listed above in more detail.
We chose to examine doxorubicin because it is the drug for
which we have the most information about specific predictions.
We chose to examine cisplatin and pemetrexed both
because the cisplatin signature contains specific genes
that have been linked to drug resistance (e.g.,
ERCC1 and ERCC4) and because these signatures are
guiding therapy now.
We chose to examine the combination validation because most cancer
patients get several drugs, not one.
We chose to examine temozolomide because
it is the most recent application
we have seen.

The goals of our reanalyses differ slightly by study, partially
reflecting differences in the types of data available.
For doxorubicin, we tried to confirm the accuracy of the test
set predictions.
For cisplatin and pemetrexed, we tried to confirm the identities of the genes
comprising the signature.
For combination therapy, we tried to clarify the
combination rules and validate predictions for the
single best performing drug (cyclophosphamide).
For temozolomide, we tried to confirm the association
between the drug named and the
results provided.

Finally, we provide a broader overview by
graphically summarizing the cell lines
designated sensitive and resistant to various drugs.

Data sources are listed in Table~\ref{tab:dataSources}.


\begin{table}
\caption{Locations of data used in our analyses. Excel files were
converted to csv files for loading into R.\break All data not posted elsewhere
is available from our web site}\label{tab:dataSources}
\begin{tabular*}{\textwidth}{@{\extracolsep{4in minus 4in}}ll@{}}
\hline
\multicolumn{2}{@{}l@{}}{\textbf{Data sources}}\\
\hline
\multicolumn{2}{@{}l@{}}{\citet{potti06} web site, accessed April 4, 2009}\\
\multicolumn{2}{@{}l@{}}{\url{http://data.genome.duke.edu/NatureMedicine.php}}\\
\quad Adria\_ALL\_data1\_n95.doc & GEO ids and Sens/Res calls for 95 samples \\
\quad Celllines\_in\_each\_predictor1.xls & Sens/Res cell lines for each drug \\
\quad ChemopredictorsParameters.xls & Binreg parameter settings for 11 drugs \\
\quad DescriptionOfPredictorGeneration.doc & Cell line selection for docetaxel \\
\quad GeneLists.zip & Probesets splitting Sens/Res cell lines\\
\quad MDACC\_data.zip & TFAC Individual and combo predictions \\
\quad ParametersForImplementingSoftware.xls & Binreg parameter settings for 7 drugs \\
\quad Binreg.zip & Metagene prediction software
\\[3pt]
\multicolumn{2}{@{}l@{}}{\citet{potti06}
web site, November 6, 2007,
 no longer posted}\\
\quad Adria\_ALL.txt & Numbers, Sens/Res labels for 144 samples, \\
  & 22 training cell lines, 122 testing samples
\\[3pt]
\multicolumn{2}{@{}l@{}}{\citet{hsu07}
journal web site, April 4, 2009}\\
\quad 10593\_Supplementary\_Information.doc & Supplementary Methods \\
\quad 10593\_Supplementary\_Table\_1.doc & Cisplatin gene list \\
\quad 10593\_Supplementary\_Table\_2.doc & Pemetrexed gene list \\
\quad Figure~1 & Cisplatin and pemetrexed heatmaps
\\[3pt]
\multicolumn{2}{@{}l@{}}{Gene Expression Omnibus (GEO), April 4, 2009}\\
\quad GSE649  & \citet{holleman04}
\quad vincristine Res \\
\quad GSE650  & \citet{holleman04}
\quad daunorubicin Sens \\
\quad GSE651  & \citet{holleman04}
\quad daunorubicin Res \\
\quad GSE2351 & \citet{lugthart05}
\quad multi-drug Res \\
\quad GSE6861 & \citet{bonnefoi07}
\quad array CEL files
\\
\hline
\end{tabular*}
\end{table}
\setcounter{table}{0}
\begin{table}
\caption{(Continued)}
\begin{tabular*}{\textwidth}{@{\extracolsep{4in minus 4in}}ll@{}}
\hline
\multicolumn{2}{@{}l@{}}{\textbf{Data sources}}\\
\hline
\multicolumn{2}{@{}l@{}}{\citet{holleman04} web site, April 4, 2009
\url{http://www.stjuderesearch.org/data/ALL4/}}\\
\quad key.xls & List of \citet{holleman04} files at GEO
\\[3pt]
\multicolumn{2}{@{}l@{}}{Drug sens., April 4, 2009,
\url{http://dtp.nci.nih.gov/docs/cancer/cancer_data.html}}\\
\quad GI50\_AUG08.BIN & GI50 values, all drugs, August 2008 release\\
\quad LC50\_AUG08.BIN & LC50 values, all drugs, August 2008 release\\
\quad TGI\_AUG08.BIN  & TGI values, all drugs, August 2008 release\\
[3pt]
\multicolumn{2}{@{}l@{}}{NCI60 array data, April 4, 2009,
\url{http://dtp.nci.nih.gov/mtargets/download.html}}\\
\quad WEB\_HOOKS\_NOV\_GC\_ALL.ZIP & MAS5.0 quantifications, U95A arrays \\
[3pt]
\multicolumn{2}{@{}l@{}}{\citet{augustine09}
\quad journal web site, April 4, 2009}\\
\quad Figure~4 & Temozolomide heatmap \\
\quad 1.pdf & Gene list and metagene scores\\
[3pt]
\multicolumn{2}{@{}l@{}}{\citet{bonnefoi07}
journal web site, April 4, 2009}\\
\quad mmc1.pdf & Webappendix with algorithm description \\
\quad mmc2.pdf & Webfigure of individual drug ROC curves \\
\quad mmc3.pdf & Webpanel listing cell lines used by drug \\
\quad mmc4.pdf & Webtable 1 listing X3p Probesets by drug \\
\quad mmc5.pdf & Webtable 2 Clinical data for samples used \\[3pt]
\multicolumn{2}{@{}l@{}}{\citet{bonnefoi07}
files obtained from
contact for GSE 6861. Not posted.}\\
\quad dataBonnefoiPaper.txt & Drug and combination scores \\
\quad HB131.CEL & CEL file missing from GEO \\
\hline
\end{tabular*}
\end{table}

\section{Case study 1: Doxorubicin}

We begin by examining the data for\break doxorubicin,
where the most 
test set prediction details
are available.\break This data was discussed by
\citet{potti06},
questioned by\break \citet{coombes07},
reaffirmed by \citet{potti07},
and revisited by \citet{potti08}
as discussed below.
In terms of documentation and reproducibility, our goal
here is to confirm the
test set prediction accuracy.

\subsection{Test data responder/nonresponder counts match
those reported}
We first acquired the raw doxorubicin (adriamycin) data (Adria\_ALL.txt)
posted by \citet{potti07}.
This file contains 144 data columns: 22 for the cell lines used as
training data, and 122 for test data
samples. Sample columns are not named, but
sensitive/resistant status is indicated for each.
A side comment notes that ``Validation data is from
GSE4698, GSE649, GSE650, GSE651, and others.''
We then checked label counts for the test data:
99 columns are labeled NR (nonresponders)
and 23 Resp (responders),
matching those reported by
\citet{potti06}.

\subsection{Training data sensitive/resistant labels are reversed}
We next tried to identify
the cell lines used in the training set by matching the numbers
reported to those in the table of NCI60 quantifications (see
Table~\ref{tab:dataSources}), focusing
on those
for the 22 cell lines used to produce the initially reported
heatmap signature [\citet{potti06},
Figure~2A].
The posted numbers have been transformed relative
to the MAS5 quantifications used earlier.
After some experimentation, we found that
the data were log-transformed, the values for each row (gene) were
centered and scaled to have mean zero and unit variance
(separately for training and test data),
exponentiated to undo the log-transform, and
rounded to two decimal places. In
order to match the numbers more precisely, we transformed
the NCI60 quantifications the same way. The initial training
data matrix for doxorubicin has 12558 rows (the 12625
Affymetrix U95Av2 probesets
less 67 controls),
but Adria\_ALL.txt has 8958. This is due to the fact that
the test data come from U133Av2 arrays, and the mapping across
platforms matched probes based on unique LocusLink IDs.
Since the rows are not labeled, we
simply tried checking all row-by-row correlations
between Adria\_ALL.txt and the transformed NCI60 data
pertaining to the 22 cell lines in the initial heatmap by brute
force. We
identified
perfect matches for all
8958, establishing that the 10 cell lines labeled ``Resistant'' are
SF-539, SNB-75, MDA-MB-435, NCI-H23, M14,
MALME-3M, SK-MEL-2, SK-MEL-28, SK-MEL-5 and UACC-62, and
the 12 cell lines labeled ``Sensitive'' are
NCI/ADR-RES, HCT-15, HT29, EKVX, NCI-H322M,
IGROV1, OVCAR-3, OVCAR-4, OVCAR-5, OVCAR-8, SK-OV-3 and
CAKI-1. These lines (in this order) do produce the heatmap shown in
Figure~2A of \citet{potti06},
and the sensitive/resistant orientation is consistent with their
Supplementary Figure~2, which notes that 
heatmaps for each predictor have
``resistant lines on the left, and sensitive on the right.''
However, the labels are reversed
relative to those 
now supplied on the
\citet{potti06}
web site in ``Cell lines used in each chemo predictor.''
Since the listing above places
NCI/ADR-RES
(``adriamycin resistant'')
in the sensitive group,
the training labels in Adria\_ALL.txt are reversed.

\begin{figure}[b]
\begin{tabular}{@{}cc@{}}
\footnotesize{(a)}&\footnotesize{(b)}\\

\includegraphics{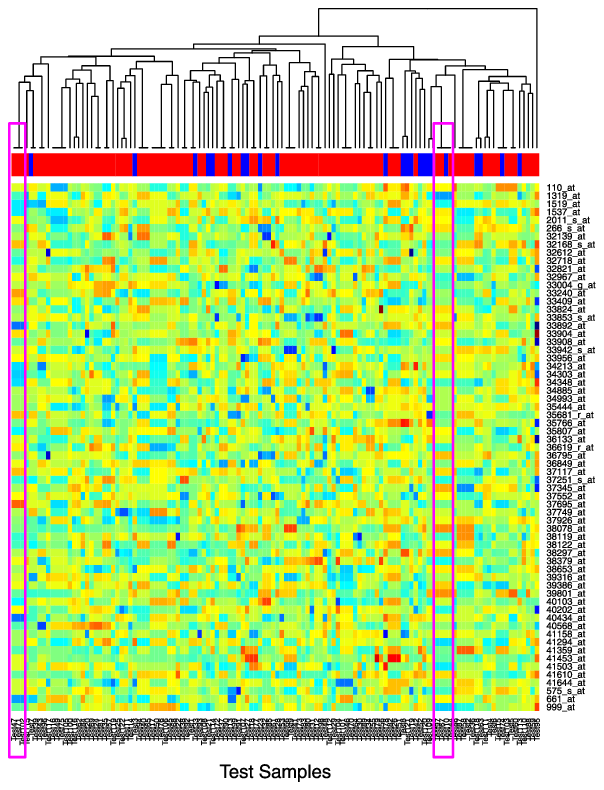}
& \includegraphics{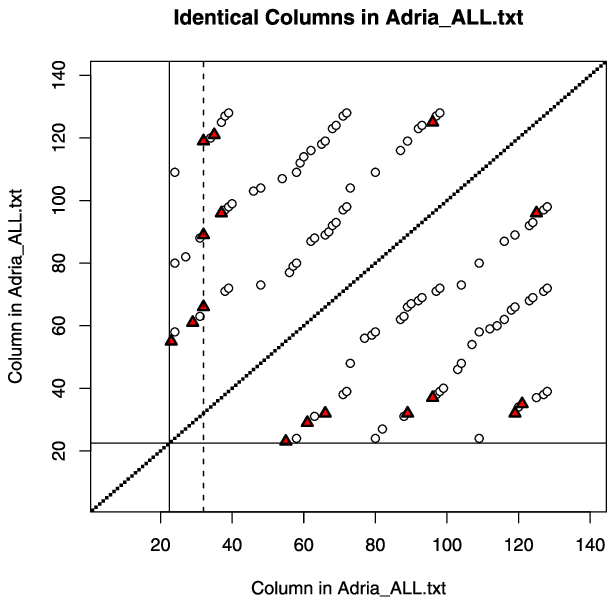}
\end{tabular}
\caption{(\textup{a}) Heatmap of test samples from
Adria\_ALL.txt, using expression values for genes
in the doxorubicin signature. Samples are labeled as \textup{``}NR\textup{''}
(red) or \textup{``}Resp\textup{''} (blue). Responders and nonresponders show
no clear separation. However, the clustering shows that some
samples are tied (horizontal segments near the
dendrogram base). Lines border one group of tied samples at
left, and another group of tied samples at right;
there are others. Colors for the tied block at right show
Resp once and NR three times.
(\textup{b}) Dotplot identifying identical columns
in the full Adria\_ALL.txt matrix. There are no unwanted
ties in the training data (lower left corner, separated by
solid lines). In the test data (upper right), however, there
are tied pairs labeled both consistently (circles)
and inconsistently (triangles). One column from
the right-hand block from (\textup{a})
is marked with a dashed line; sample
32 (main diagonal) is labeled \textup{``}Resp,\textup{''} but samples 66, 89,
and 117 are labeled \textup{``}NR\textup{''} (triangles).}\label{fig:doxorubicinFigure}
\end{figure}

\subsection{Heatmaps show sample duplication
in the test data [Figure~\protect\ref{fig:doxorubicinFigure}(\textup{a})]}

In order to see how well the doxorubicin signature
separates test set responders from nonresponders,
we extracted the test set expression values
for the probesets in the doxorubicin signature
(using the identifications from the previous step),
log-transformed them, and clustered the samples
[Figure~\ref{fig:doxorubicinFigure}(a)].
We did not see clear separation of responders
from nonresponders, but we did see
``blocks'' of samples having exactly the same profiles: some
test data samples were reused. Two such blocks are
marked.

\subsection{Only 84$/$122 test samples are distinct; some
samples are labeled both sensitive and resistant [Figure~\protect\ref{fig:doxorubicinFigure}(\textup{b})]}

In order to get a more precise idea of the extent
of the duplication, we examined all pairwise sample correlations
for Adria\_ALL.txt. Figure~\ref{fig:doxorubicinFigure}(b)
highlights pairs with correlations
above 0.9999 (duplicates). Circles indicate
duplicates with consistent labels; triangles indicate duplicates
where the same sample was labeled sensitive in one
column and resistant in the other. Only 84 of the 122
test samples are distinct:
60 are present once,
14~twice,
6 three times, and
4 four times.
Columns 32 (dashed line), 66, 89 and 117 in Adria\_ALL.txt
are all the same, but they are labeled
Sensitive, Resistant, Resistant and Resistant respectively
[this is the set of samples marked at the right
in Figure~\ref{fig:doxorubicinFigure}(a)].

\subsection{Communication with the journal elicited a second correction}

This correction [\citet{potti08}]
notes that
\begin{quote}
``of the 122 samples assayed for sensitivity to daunorubicin
for which the authors applied a predictor of adriamycin
sensitivity, 27 samples were replicated owing to the fact
that the same samples were included in several separate
series files in the Gene Expression Omnibus generated in
2004 and 2005, which were the source of the data provided
for the study.''
\end{quote}
This correction also notes that data were acquired from
two other sources in addition to GSE650 and GSE651:
GSE2351 [\citet{lugthart05}]
and GSE649
[\citet{holleman04}].

\begin{table}[b]
\caption{The first 20 rows of Adria\_ALL\_data1\_n95.doc.
Visual examination shows rows 3/4, 9/10 and 17/18 (marked)
are the same.
Rows 17/18 label the same sample both ways}\label{tab:adria95First20}
\begin{tabular*}{\textwidth}{@{\extracolsep{4in minus 4in}}d{2.0}ccccc@{}}
\hline
\multicolumn{1}{@{}l}{\textbf{Row}} & \textbf{Sample ID} & \textbf{Response} & \textbf{Row} & \textbf{Sample ID} & \textbf{Response} \\
\hline
 1 &GSM44303   &RES       &11 &GSM9694   &RES\\
 2 &GSM44304   &RES       &12 &GSM9695   &RES\\
 3 &GSM9653*    &RES       &13 &GSM9696   &RES\\
 4 &GSM9653*    &RES       &14 &GSM9698   &RES\\
 5 &GSM9654    &RES       &15 &GSM9699   &SEN\\
 6 &GSM9655    &RES       &16 &GSM9701   &RES\\
 7 &GSM9656    &RES       &17 &GSM9708*   &RES\\
 8 &GSM9657    &RES       &18 &GSM9708*   &SEN\\
 9 &GSM9658*    &SEN       &19 &GSM9709   &RES\\
10 &GSM9658*    &SEN       &20 &GSM9711   &RES\\
\hline
\end{tabular*}
\end{table}

\subsection{The new data also has duplications and samples listed both
ways (Table~\protect\ref{tab:adria95First20})}

At the time of the \citet{potti08}
correction, Adria\_ALL.txt was removed
from the
web site and replaced with Adria\_ALL\_data1\_n95.doc.
Adria\_ALL\_data1\_n95.doc gives no quantifications,
but rather lists 95 GEO array ids
and gives a sensitive/resistant label for each.
The first twenty data rows from this file are shown
in Table~\ref{tab:adria95First20}.
Rows 3/4, 9/10 and 17/18 list the same samples.
Further, the status labels in rows 17/18 conflict.
Looking at all the names shows that
only 80 are distinct: 15 are duplicated,
and 6 of these show the same sample labeled both RES and SEN.


\subsection{At least 3$/$8 of the test data is incorrectly
labeled resistant (Table~\protect\ref{tab:crossTab})}
Given the test sample ids,
it is also possible to compare the classifications
\citet{potti08}
assigned (not \textit{predicted},
but rather what they are trying to predict)
to the 80 unique samples
with the classifications assigned using the rules
from \citet{holleman04},
the source of
the test data
(Table~\ref{tab:crossTab}).
Between 29 and 35 samples
\citet{holleman04}
would call sensitive are classed by
\citet{potti08}
as resistant, with the uncertainty
due to
duplicate samples with inconsistent calls.
There are 10 samples
\citet{holleman04}
would call ``intermediate'' that are classed by
\citet{potti08}
as resistant. All of the changes add to the number of ``resistant'' cases.
This shift is not achievable by simply moving the LC50 cutoff to
redefine the sensitive/resistant boundary,
as LC50 values for ``sensitive'' and ``resistant'' samples overlap.


\begin{table}
\caption{Joint classifications of the 80 distinct samples
in Adria\_ALL\_data1\_n95.doc.
Holleman et~al. (\protect\citeyear{holleman04})
class most
samples as sensitive to daunorubicin;
Potti et~al. (\protect\citeyear{potti06})
class most as resistant.
Adria\_ALL\_data1\_n95.doc
has 95 rows with 15 duplicates; 6 duplicates
are labeled inconsistently and are classed as \textup{``}Both\textup{''}}\label{tab:crossTab}
\begin{tabular*}{\textwidth}{@{\extracolsep{4in minus 4in}}lcd{2.0}d{2.0}d{2.0}r@{}}
\hline
\multicolumn{2}{c}{} & \multicolumn{3}{c}{\textbf{\citet{holleman04}
    classifications}} & \\[-6pt]
    \multicolumn{2}{c}{} & \multicolumn{3}{c}{\hrulefill}&\\
\multicolumn{2}{c}{}
& \multicolumn{1}{c}{\textbf{Sensitive}}
& \multicolumn{1}{c}{\textbf{Intermediate}}
& \multicolumn{1}{c}{\textbf{Resistant}} & \\
\hline
\citet{potti06} 
                & Sensitive & 13 &  0 &  0 & 13\\
\quad classifications & Resistant & 29 & 10 & 22 & 61 \\
                & Both      &  6 &  0 &  0 & 6\\
\multicolumn{2}{c}{}        & 48 & 10 & 22 & \\
\hline
\end{tabular*}
\end{table}

\subsection{Summary}
Poor documentation hid both sensitive/resistant label
reversal, and the incorrect use of duplicate (and in
some cases mislabeled) samples. These problems were
hidden well enough to survive two explicit corrections.
Code and documentation for this case study
are given in Supplementary File~1 [\citet{baggerly09a}].

\section{Case study 2: Cisplatin and pemetrexed}

We next examined data for cisplatin and pemetrexed, where (a) specific
genes in the cisplatin signature (ERCC1, ERCC4) were identified as providing
a plausible rationale for its effectiveness and (b) the
signatures are
guiding therapy.
Most signatures we discuss (including that for pemetrexed)
are assembled from the NCI60, but
\citet{hsu07}
comment that

\begin{quote}
``The collection of data in the NCI-60 data occasionally does not represent
a significant diversity in resistant and sensitive cell lines to any given drug.
Thus, if a drug screening experiment did not result in widely variable GI50/IC50
and/or LC50 data, the generation of a genomic predictor is not possible using
our methods, as in the case of cisplatin.''
\end{quote}
Thus,
\citet{hsu07}
assembled the cisplatin signature
from a panel of 30 cell lines profiled by
\citet{gyorffy06}.
Gy\"{o}rffy et al.~supply both U133A array quantifications
and classifications
of which cell lines were sensitive, intermediate or resistant
in their response to various drugs, including cisplatin (their
Figure~2).
In terms of documentation and reproducibility, our goal
here is to recreate the signature.
We acquired the cisplatin and pemetrexed gene lists from
\citet{hsu07}'s
Supplementary Tables~1 and~2, respectively.
We acquired the
\citet{gyorffy06}
expression data from their supporting Table~1.
We acquired the \mbox{\textttm{binreg}} Matlab scripts
used for model fitting and heatmap generation
from the
\citet{potti06}
web site.
As the analyses are largely parallel, we focus on
cisplatin first, turning to pemetrexed only in the
next to last subsection.

\begin{figure}
\begin{tabular}{@{}c@{\hspace*{7pt}}c@{}}
\footnotesize{(a)}&\footnotesize{(b)}\\

\includegraphics{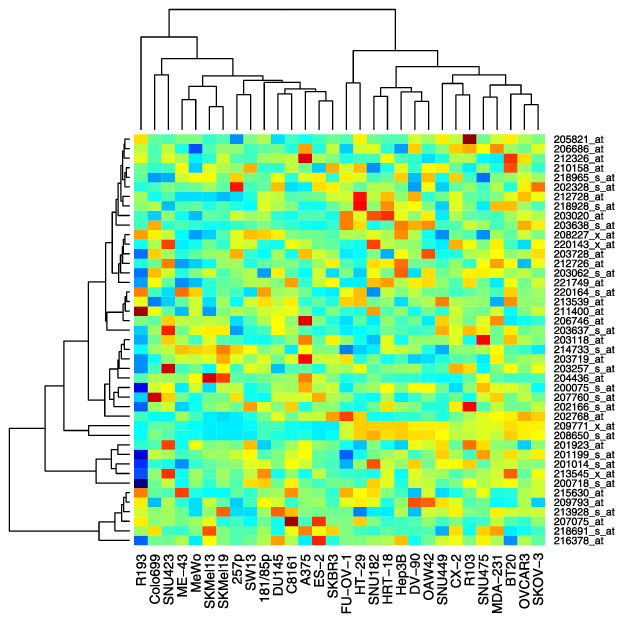}
&\includegraphics{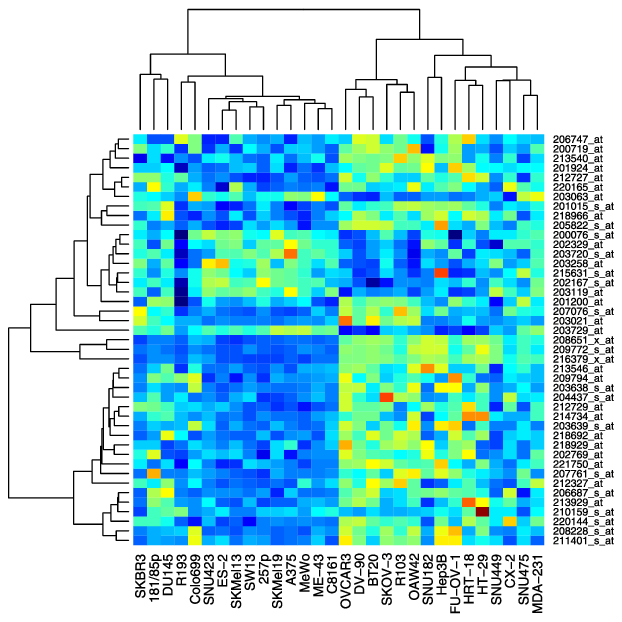}\\[3pt]
\footnotesize{(c)}&\footnotesize{(d)}\\

\includegraphics{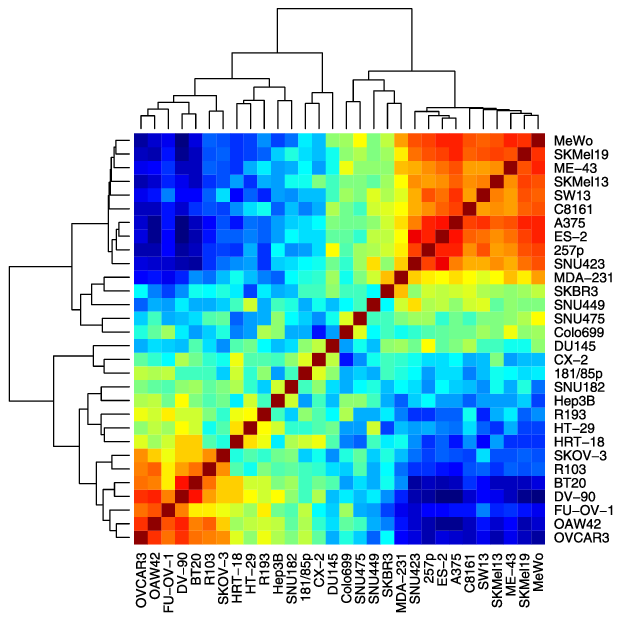}
&\includegraphics{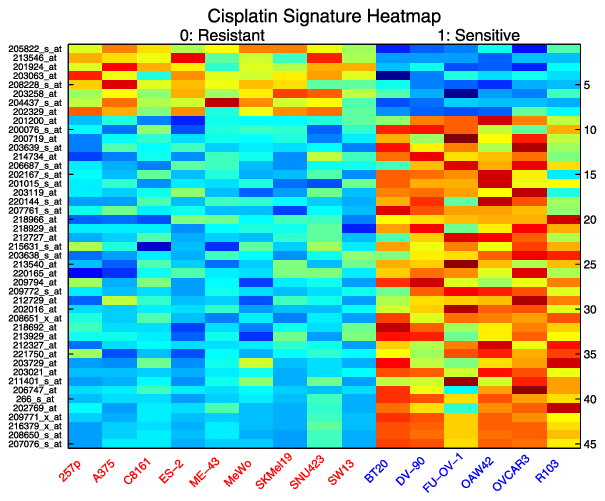}\\
\end{tabular}
\caption{Images used in reconstructing the Hsu et~al. (\protect\citeyear{hsu07})
heatmap for cisplatin. (\textup{a}) Heatmap
of the named cisplatin signature genes across the 30
Gy\"{o}rffy et~al. (\protect\citeyear{gyorffy06})
cell lines. There is very
little structure visible. (\textup{b}) Heatmap
across the same cell lines using expression values for
genes obtained by \textup{``}offsetting\textup{''} by one row (e.g., replacing
200075\_s\_at with 200076\_at). There is clear separating
structure.
(\textup{c}) Heatmap clustering pairwise sample correlations
from the expression submatrix shown in (\textup{b}). Red clusters
(high correlations) in the lower left and upper right suggest
initial guesses at the lines that should be treated as
\textup{``}sensitive\textup{''} and \textup{``}resistant\textup{''} respectively. (\textup{d}) Heatmap
produced by \texttt{binreg} using cell lines noted in the text.
This is an exact match for the heatmap
in Hsu et~al. (\protect\citeyear{hsu07}),
confirming the cell lines and genes involved.}\label{fig:cisplatinFigure}
\end{figure}

\subsection{A heatmap using the cisplatin
genes shows no separation of the cell
lines [Figure~\protect\ref{fig:cisplatinFigure}(\textup{a})]}
In order to see how clearly the sensitive and resistant cisplatin cell lines
were separated, we extracted and clustered the expression submatrix for the
signature genes across all 30 cell lines
[Figure~\ref{fig:cisplatinFigure}(a)].
The heatmap shows no clear split between sensitive
and resistant cell lines.


\subsection{A heatmap using offset cisplatin
genes shows clear separation of the cell
lines [Figure~\protect\ref{fig:cisplatinFigure}(\textup{b})]}
\citet{coombes07}
noted that several of the
gene lists initially reported by
\citet{potti06}
were ``off-by-one'' due to an indexing error, so
we also extracted and clustered the expression submatrix for
the ``offset'' cisplatin signature genes across all 30 cell lines,
producing the heatmap shown in
Figure~\ref{fig:cisplatinFigure}(b).
This offset involves a single row shift:
for example, quantifications from row 98 (probeset 200076\_at)
of the
\citet{gyorffy06}
table were
used instead of those from row 97 (probeset 200075\_s\_at).
This heatmap shows a clear split between sensitive
and resistant cell lines.

\subsection{Clustering correlations suggests the cell lines involved
[Figure~\protect\ref{fig:cisplatinFigure}(\textup{c})]}
\citet{gyorffy06}
list 20 and 7 lines as
resistant and sensitive to cisplatin, respectively (3 are listed
as intermediate), but
\citet{hsu07}
show 9 and 6. In order to figure out
which cell lines were used, we clustered the correlations between
sample columns from the expression submatrix for the offset genes
shown in Figure~\ref{fig:cisplatinFigure}(b).
This clustering, shown in Figure~\ref{fig:cisplatinFigure}(c),
shows two clear groups of cell lines: 10 in the upper right
and 7 in the lower left.
We used the sensitive and resistant
labels from
\citet{gyorffy06}
to suggest which
group was which (upper right is resistant).
Checking the labels assigned by
\citet{gyorffy06}
shows that SKMel13 (Sensitive) and Skov3 (Intermediate) do not
fit with the others of their respective groups. Omitting these
gives the following 9 ``Resistant'' cell lines:
257P, A375, C8161, ES2, me43, MeWo,
SKMel19, SNU423 and Sw13; and 6 ``Sensitive'' cell lines:
BT20, DV90, FUOV1, OAW42, OVKAR and R103.
We checked this further using a brute force ``steepest ascent''
method with their \textttm{binreg} software as follows.
Starting with our initial guess
of 10 and 7 lines, we labeled each of the 30
cell lines
as resistant, sensitive or not used. We gave our guess
a score equal to the number of offset probeset ids matching the
\textttm{binreg} output
(our starting score was 30).
We then examined all 60 combinations of cell lines that could be
reached by changing the status of one cell line from its current
state to resistant, sensitive or not used, and moved to the
combination with the highest score (36, obtained by dropping
SKOV-3 from the sensitive group), and iterating
until a local maximum was reached (41, on the second step,
after dropping SKMel-13 from the resistant group).

\subsection{\texorpdfstring{Applying \texttt{binreg} perfectly reproduces the reported
heatmap [Fig-\break ure~\protect\ref{fig:cisplatinFigure}(\textup{d})]}{Applying \texttt{binreg} perfectly
reproduces the reported heatmap [Figure~\protect\ref{fig:cisplatinFigure}(\textup{d})]}}

When we apply \textttm{binreg} to identify the top 45
genes differentiating the cell lines named above,
\textttm{binreg} produces
the heatmap shown in Figure~\ref{fig:cisplatinFigure}(d).
This is an exact match to the cisplatin heatmap reported by
\citet{hsu07},
confirming the cell lines
used and identifying the genes involved.

\subsection{The software produces 41$/$45 offset genes; the others are
the ones explicitly mentioned}
We compared the cisplatin gene list \textttm{binreg} produced for the
matching heatmap with the gene list
\citet{hsu07}
reported. We match 41 of the 45~probesets using
the offset gene list.
The four probesets we do not match after offsetting (the ``outliers'')
are 203719\_at (ERCC1), 210158\_at (ERCC4),
228131\_at (ERCC1) and 231971\_at (FANCM, associated with DNA Repair),
which \citet{hsu07}
explicitly mention as interesting components
of the signature.

\subsection{Two of the \textup{``}outlier\textup{''} genes are not on the arrays used}

It is actually vacuous to say we cannot match 228131\_at (ERCC1)
and 231971\_at (FANCM; DNA Repair), as
\citet{gyorffy06}
report no quantifications for these probesets. Checking the
U133 probeset annotation files available from Affymetrix
(\url{http://www.affymetrix.com}) shows
that these probesets are on the U133B array platform, not the U133A,
so
\citet{gyorffy06}
did not measure them.
[The heatmaps in Figures~\ref{fig:cisplatinFigure}(a) and (b) only have 43 rows.]

\subsection{ERCC1 and ERCC4 have been outliers before}
Coombes, Wang and Baggerly (\citeyear{coombes07})
noted that several 
gene lists initially
reported by
\citet{potti06}
contained genes
that \textttm{binreg} did not produce.
ERCC1 and/or ERCC4 were outliers in the
gene lists initially reported for docetaxel, paclitaxel and
doxorubicin as well as cisplatin. The other outliers
are shown in Table~\ref{tab:Outliers}.


\begin{table}
\caption{\textup{``}Outlier\textup{''} probesets in the initially reported gene lists
for
Potti et~al. (\protect\citeyear{potti06})
(docetaxel, paclitaxel, doxorubicin)
and
Hsu et~al. (\protect\citeyear{hsu07})
(cisplatin). ERCC1 and ERCC4
are common. Probesets for cisplatin marked with an asterisk
come from the U133B array platform, not the U133As used.\break
The table does not include 14 outliers in the initial
signature for docetaxel that are also found in the list
reported by
Chang et~al. (\protect\citeyear{chang03})
as being effective
separators in the docetaxel test set}\label{tab:Outliers}
\begin{tabular*}{\textwidth}{@{\extracolsep{4in minus 4in}}lccccc@{}}
\hline
\textbf{U95Av2 PS} & \textbf{Docetaxel} & \textbf{Paclitaxel} & \textbf{Doxorubicin} & \textbf{U133A PS} & \textbf{Cisplatin} \\
\hline
114\_r\_at & & MAPT && 203719\_at & ERCC1 \\
1258\_s\_at & ERCC4 & ERCC4 & ERCC4 & 210158\_at &  ERCC4 \\
1802\_s\_at & ERBB2 & ERBB2 && 228131\_at & ERCC1* \\
1847\_s\_at & & & BCL2 & 231971\_at & FANCM* \\
1878\_g\_at & ERCC1 & ERCC1 &&& \\
1909\_at & & & BCL2 && \\
1910\_s\_at & & & BCL2 && \\
2034\_s\_at & & & CDKN1B && \\
33047\_at   & BCL2L11 & BCL2L11&&& \\
36519\_at & & ERCC1 &&& \\
40567\_at   & K-ALPHA-1 & K-ALPHA-1 & & & \\
\hline
\end{tabular*}
\end{table}

\subsection{Genes are offset, and sensitive/resistant labels are reversed for pemetrexed}
Having matched cisplatin, we then turned to pemetrexed.
We
used the
steepest ascent method described above
to identify the cell lines and genes in the pemetrexed
signature. After offsetting, we are able to
match all 85 genes and to perfectly match the heatmap shown. The
8 ``Resistant'' cell lines are
K-562, MOLT-4, HL-60(TB), MCF7, HCC-2998, HCT-116,
NCI-H460 and TK-10, and the 10 ``Sensitive'' cell lines are
SNB-19, HS 578T, MDA-MB-231/ATCC, MDA-MB-435,
NCI-H226, M14, MALME-3M, SK-MEL-2, SK-MEL-28 and SN12C.
Unfortunately, checking the GI50 data for pemetrexed shows
that the sensitive/resistant labels are reversed.

\subsection{Summary}
Poor documentation hid an off-by-one indexing error
affecting all genes reported, the inclusion of genes
from other sources, including other arrays (the outliers),
and a sensitive/resistant label reversal.
Our analyses of these signatures are contained in
Supplementary File~2 [\citet{baggerly09b}].

\section{Case study 3: Combination therapy}

We next examined the approaches used to investigate
combination therapy.
\citet{potti06}
mention success in deriving predictions for breast cancer patients
treated with a combination of paclitaxel (taxol), 5-fluorouracil,
adriamycin and cyclophosphamide (TFAC).
Their methods note that the combination predictions were derived
from those for the individual drugs
using ``the theorem for combined probabilities as described by
William Feller.''
Later,
\citet{bonnefoi07}
provided a ``validation''
of the combination approach, using it to predict patient response
to two alternative therapies: taxotere (docetaxel) followed by epirubicin
(similar to doxorubicin) and taxotere (TET), and
fluorouracil, epirubicin and cyclophosphamide (FEC).
In terms of documentation and reproducibility, our goals
here are to clarify the combination rules and to check
predictions for the best single drug.
We acquired raw CEL files and array quantifications
from GEO (GSE6861; posted quantifications have been revised
since we obtained them at the end of 2007). We obtained
a missing CEL file (HB131) and
a table giving the individual and combination drug
predictions from the statistical team in Lausanne which
used the predictions to construct ROC curves.

\subsection{Treatment is confounded with run date}
\begin{figure}
\begin{tabular}{@{}c@{\hspace*{7pt}}c@{}}
\footnotesize{(a)}&\footnotesize{(b)}\\

\includegraphics{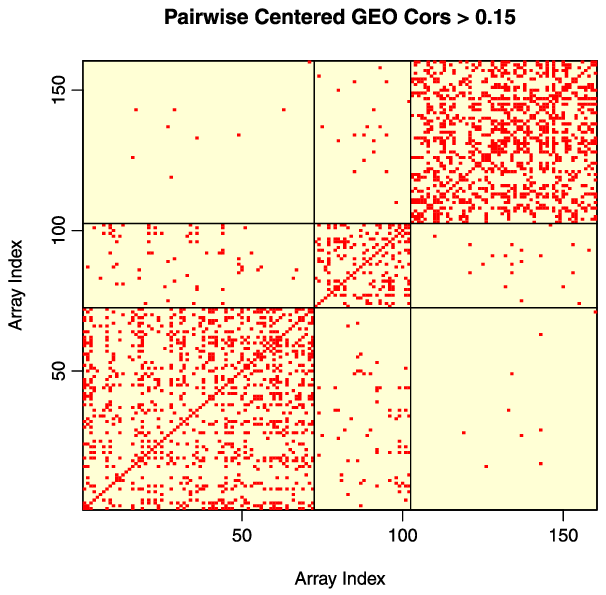}
&\includegraphics{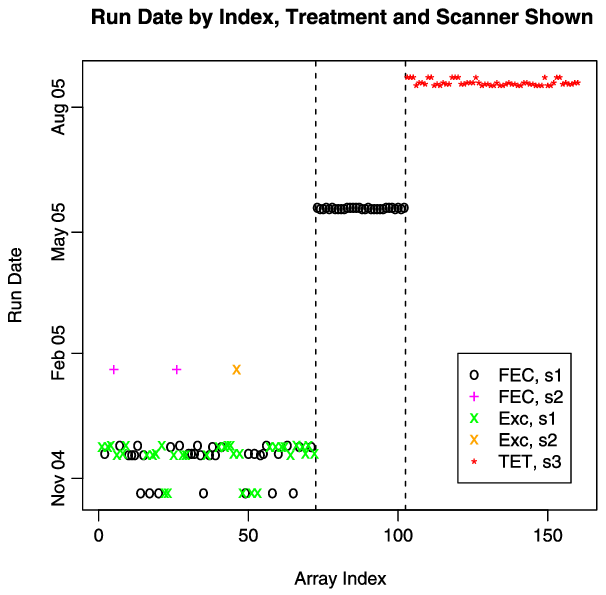}\\
\footnotesize{(c)}&\footnotesize{(d)}\\

\includegraphics{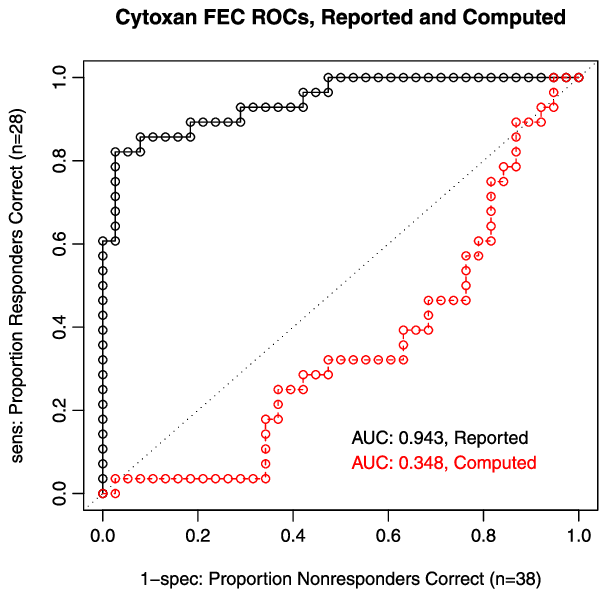}
&\includegraphics{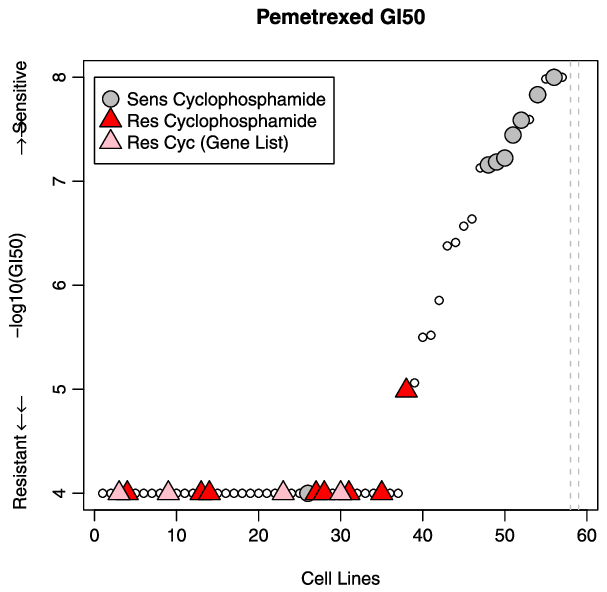}\\
\end{tabular}
\caption{Aspects of the Bonnefoi et~al. (\protect\citeyear{bonnefoi07})
study of combination therapy. (\textup{a}) Examining array quantifications
for high pairwise correlations shows three blocks in the data.
(\textup{b}) Plotting array run date by index shows the same three blocks.
The first block contains half of the patients treated with FEC;
the other samples in this block were excluded. The second block
contains the other half of the patients treated with FEC. The
third block contains all of the patients treated with TET, with
the arrays run on a different scanner.
(\textup{c}) ROC curves for the single best set of drug predictions:
cyclophosphamide for FEC. The reported curve has an AUC of 0.943,
indicating extremely good prediction. Our best approximation
has an AUC of 0.348, indicating if anything performance worse
than chance.
(\textup{d}) Drug sensitivity data for pemetrexed, with the cell
lines named for cyclophosphamide indicated. The corresponding
plot for cyclophosphamide itself is roughly flat.
Pink triangles indicate cell lines that were not reported
but are required to generate the gene list reported.}\label{fig:bonnefoiFigure}
\end{figure}

We first checked for gross differences in the
array data. Examining high pairwise
correlations shows the presence
of three clear blocks
[Figure~\ref{fig:bonnefoiFigure}(a)]. We then
extracted run date and lab information from the
CEL file headers and plotted the data by run
date [Figure~\ref{fig:bonnefoiFigure}(b)];
the three clusters correspond to three
major run blocks. Different symbols in
Figure~\ref{fig:bonnefoiFigure}(b) show that the first block contains
half of the patients treated with FEC, and that
roughly half of the samples run at this time were
excluded from the final analysis. The second
block contains the second half of the patients
treated with FEC. The third block contains
all of the patients treated with TET; all of these
arrays were run on a different scanner than those
from the first two blocks. There is
perfect confounding of run date with treatment.


\subsection{Three different combination rules were used}
We then tried to identify the rules used to combine
individual drug predictions into a combination score.
Letting $P$($\cdot$) indicate probability of sensitivity,
the rules used are as follows:
\begin{eqnarray*}
  P(TFAC) &=& P(T) + P(F) + P(A) + P(C) - P(T)P(F)P(A)P(C),\\
  P(TET) &=& P(ET) = \max[P(E),P(T)],
\end{eqnarray*}
and
\begin{eqnarray*}
  P(FEC) = \tfrac{5}{8}[P(F) + P(E) + P(C)] - \tfrac{1}{4}.
\end{eqnarray*}
Values obtained with the first rule were
all greater than 1, so the largest value was set to 1,
the smallest to 0, and the others fit by linear
interpolation. Values bigger than 1 or less than 0
using the third rule were set to 1 and 0, respectively.
These rules are not explicitly stated in the methods;
we inferred them either from formulae embedded in
Excel files (TFAC) or from exploratory data analysis
(TET and FEC). None of these rules are standard.
Since all three rules are different, it is
not clear what rule was validated.

\subsection{We can\textup{'}t match the accuracy for the best drug/treatment
combination [Figure~\protect\ref{fig:bonnefoiFigure}(\textup{c})]}
We also tried to roughly replicate the sensitivity
predictions reported for the best performing component:
cyclyophosphamide in the FEC group. To do this, we first
matched the X3P platform probesets
\citet{bonnefoi07}
report for~cyclophosphamide with the U95Av2 platform
probesets now available as part of the supplementary
information for
\citet{potti06}.
This was
accomplished first by using ``chip comparer''
(\url{http://chipcomparer.genome.duke.edu}), 
which defined one-to-one mappings for most
of the probesets, and using the platform annotation
available from either GEO or GeneCards to resolve
any ambiguities. We then approximated the \textttm{binreg} metagene
(SVD) computation
in R code,
and produced ROC curves using both the reported
predictions and those we assembled [Figure~\ref{fig:bonnefoiFigure}(c)].
The ROC curve
assembled with the reported predictions
(AUC of 0.943) perfectly matches that given by
\citet{bonnefoi07}.
Our own curve (AUC of 0.348) is qualitatively
different.

\subsection{Sensitivity to cyclophosphamide doesn\textup{'}t separate
the cell lines used. Sensitivity to pemetrexed does}
We also examined the drug sensitivity data for cyclophosphamide
(NSC 26271) to clarify how the cell lines were chosen, but saw
no differential activity. This is driven by the fact that
cyclophosphamide is a prodrug (a drug that needs to be
processed by the body to produce the active form), and
has no direct effect on cell lines.
The cell lines used for cyclophosphamide match those
found above in the signature for pemetrexed;
sensitivity data for pemetrexed is shown in Figure~\ref{fig:bonnefoiFigure}(d).

\subsection{Summary}
Design confounding was not mentioned. Poor documentation
obscured the fact that different combination rules were
used, and leaves both the computation of scores and selection
of cell lines for cyclophosphamide unclear.
Details of our analyses are given in Supplementary
File~3 [\citet{baggerly09c}].

\section{Case study 4: Temozolomide}

We next examined the signature for temozolomide
reported by
\citet{augustine09},
who discuss how it might be useful for predicting
response in melanoma. The initial
signature involved 45 genes separating 9 resistant
from 6 sensitive cell lines, with cell lines coming
from the NCI-60 panel.
In terms of documentation and reproducibility, our goal
here is to match the
results
to the drug named.
We acquired the heatmap from
\citet{augustine09}'s
Figure~4A
and the genes from their supplementary
table.

\subsection{The gene list doesn\textup{'}t match the heatmap}
We first tried to confirm the behavior
of the individual genes. The gene list indicates that
8 probesets are more strongly expressed in the
resistant lines, with the other 37 more strongly
expressed in the sensitive lines. However,
three genes (RRAGD, SFN and SLC43A3) are
listed as higher in both groups (these genes were
interrogated by multiple probesets).

\begin{figure}
\begin{tabular}{@{}cc@{}}
\footnotesize{(a)}&\footnotesize{(b)}\\

\includegraphics{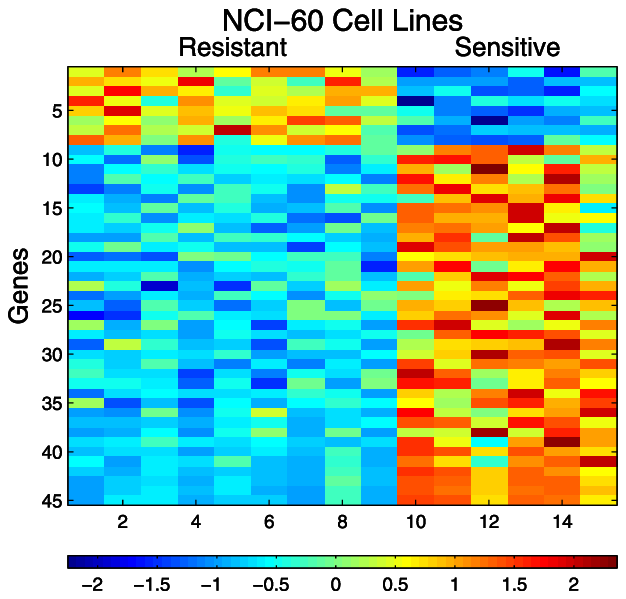}
&\includegraphics{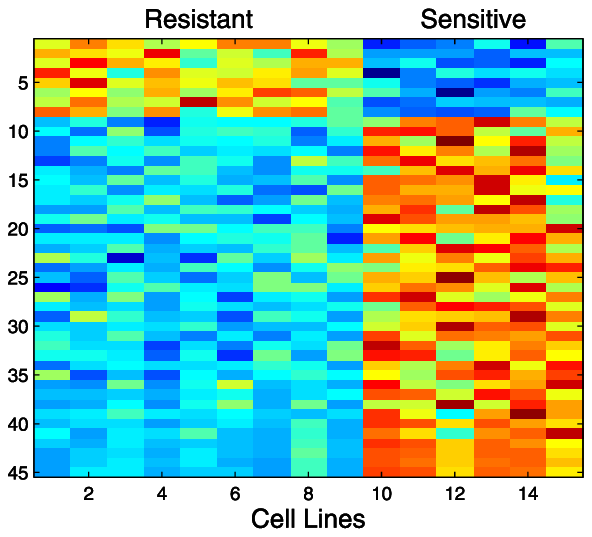}
\end{tabular}
\caption{Approximations to (\textup{a}) the heatmap initially
presented in Figure~4\textup{A} of Augustine et~al. (\protect\citeyear{augustine09})
for temozolomide, with lines reportedly chosen from the
NCI-60 cell line panel, and (\textup{b}) the heatmap presented
in Figure~1 of Hsu et~al. (\protect\citeyear{hsu07})
for cisplatin,
with cell lines chosen from the 30-line panel of
Gy\"{o}rffy et al. (\protect\citeyear{gyorffy06}).
The heatmaps
are the same. We have independently generated
the cisplatin heatmap using the Gy\"{o}rffy et al. (\protect\citeyear{gyorffy06})
data,
but the temozolomide heatmap is neither for temozolomide
nor from the NCI-60 panel.}\label{fig:temozolomideFigure}
\vspace*{-3pt}
\end{figure}

\subsection{The heatmap is that published for cisplatin
(Figure~\protect\ref{fig:temozolomideFigure})}
We tried to match the heatmap reported
following the approach described above for cisplatin,
but were unsuccessful.
We then visually compared the temozolomide and
cisplatin heatmaps (approximated in Figure~\ref{fig:temozolomideFigure}).
The heatmaps are identical.
Since we independently regenerated the cisplatin
heatmap using cell lines from
\citet{gyorffy06},
this heatmap does not correspond to temozolomide and
was not derived from the NCI-60 cell lines.
Since the reported gene list does not match the list for
cisplatin (even allowing for offsetting), we presumed
that it should correspond to the true list for temozolomide.


\subsection{Journal communication
led to a new heatmap with different problems}
In correction,
a new figure was supplied [\citet{augustine09}].
The corrected heatmap involves 150 genes, not 45, so
the initial gene list evidently matches neither cisplatin
or temzolomide. The revised heatmap
shows 5 resistant lines, 5 sensitive lines and 150 probesets,
as opposed to 9, 6 and 45 before. The fraction of probesets
higher in the resistant group has changed from 8$/$45 to about
110$/$150. As noted in the \hyperref[sec1]{Introduction},
the initial heatmap was described in detail
in the text of
\citet{augustine09},
but none of these differences were noted.
The new caption names 93 genes as higher in the resistant group.
Visual inspection shows
about 110.

\subsection{Summary}
Poor documentation led a report on drug A to include
a heatmap for drug B and a gene list for drug C.
These results are based on simple visual
inspection and counting, and are not documented further.

\section{Case study 5: Surveying cell lines used}

In light of the above issues, we decided to assemble
a more extensive overview of
which cell lines
were treated as sensitive or resistant for various drugs.
We considered 12 sources of information about the
signatures for 10 drugs: docetaxel (D), paclitaxel (P),
doxorubicin (adriamycin, A), fluorouracil (F),
topotecan (T), etoposide (E), cyclophosphamide~(C),
pemetrexed (Pem), cisplatin (Cis) and temozolomide (Tem).
The sources are as follows:
\begin{enumerate}[11.]
  \item[1.] Heatmaps from \citet{potti06}
    (D, P, A, F, T, E), November 2006.
  \item[2.] Heatmaps from \citet{hsu07}
    (Cis, Pem), October 2007.
  \item[3.] Gene lists from \citet{potti07}
    (D, P, A, F, T, E, C), November 2007.*
  \item[4.] Docetaxel quantifications from the \citet{potti06}
    website
    (D), November 2007.
  \item[5.] Doxorubicin quantifications from the \citet{potti06}
    website
    (A), November 2007.
  \item[6.] The ``list of cell lines used'' and the ``description of
    predictor generation'' from the \citet{potti06}
    website
    (D, P, A, F, T, E, C), November 2007.*
  \item[7.] The ``list of cell lines used'' supplied as a webpanel
    supplement to\break \citet{bonnefoi07}
    (D, A, F, C), December 2007.
  \item[8.] Numbers of sensitive and resistant lines used by
    \citet{salter08}
    (P, A, F,~C), April 2008.*
  \item[9.] The ``list of cell lines used''
    posted on the \citet{potti06}
    website
    (D, P, A, F, T, E, C), August 2008.
  \item[10.] The ``list of cell lines used'' by \citet{bonnefoi07}
    after correction (D, A, F, C), September 2008.
  \item[11.] Numbers of sensitive and resistant lines named by
    \citet{riedel08}
    (D, P, A, F, T, E, C), October 2008.*
  \item[12.] Heatmap from \citet{augustine09}
    (Tem), January 2009.
\end{enumerate}
We draw inferences from heatmaps and gene lists when we are
able to exactly match the reported results with the \textttm{binreg}
software used by
\citet{potti06}.
This uniquely identifies the
two groups being contrasted, but not the direction (which group is sensitive).
Direction
is inferred from other statements
in the relevant papers about what the figures represent.
In some cases (*'s above) either the
direction or the identity of the cell lines is not precisely
specified, so some information must be inferred from other sources.
These ``other sources'' need not be complex; in the case of
\citet{salter08},
for example, we simply examined
the heatmaps and counted the numbers of cell lines on the left
and right of the major divide.

\begin{figure}

\includegraphics{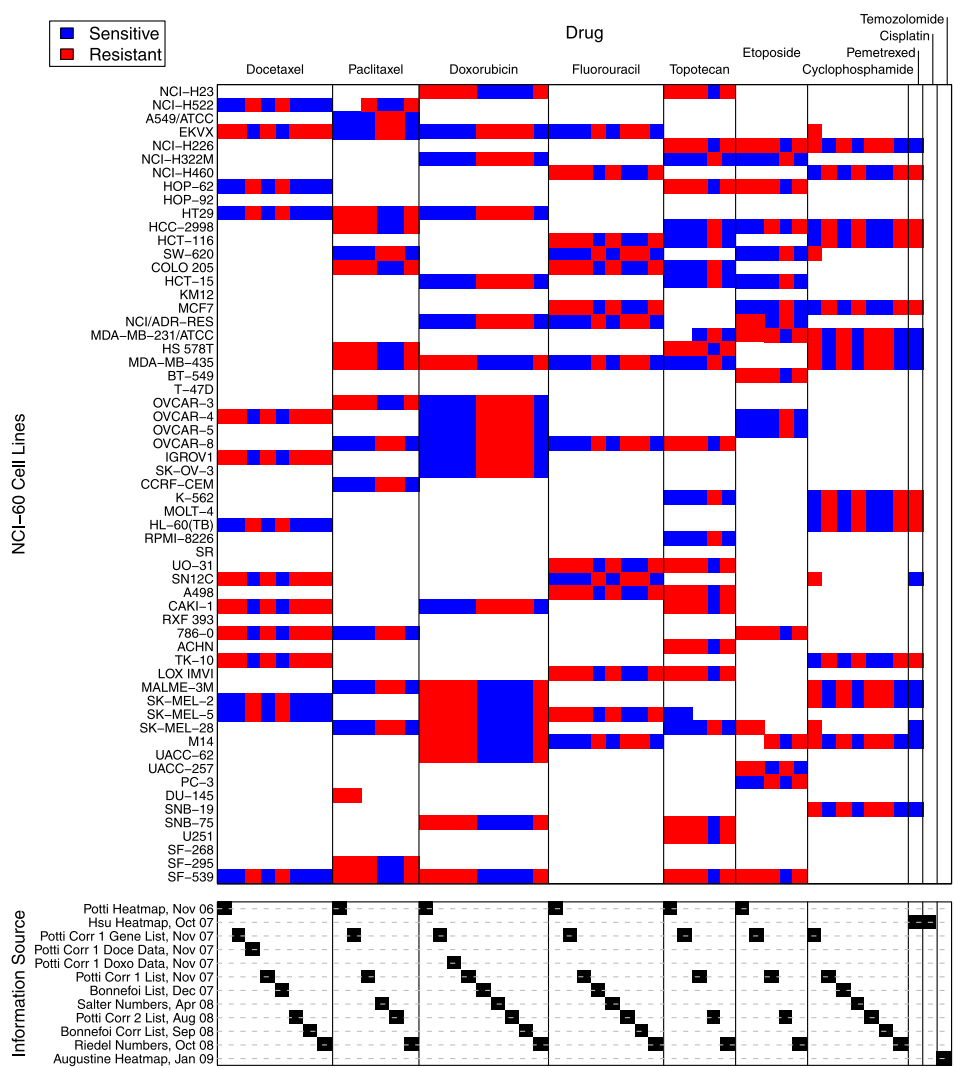}

\caption{Mapping
of which NCI-60 cell lines were
used as sensitive or resistant
for which drugs, and
the information source used for the inference. For example,
there are 8 sources of information (columns) for docetaxel; in column 3,
corresponding to the docetaxel quantifications supplied in November 2007,
cell line NCI-H522 was listed as resistant. All drug groupings with
more than one column show color flips, corresponding to reversal
of the sensitive/resistant labeling. Cell lines
are different for most drugs save cyclophosphamide/pemetrexed.
No cell lines are indicated
for cisplatin or temozolomide; the first
signature was derived from a different set of cell lines, and
in the heatmap reported for temozolomide
is actually the heatmap for cisplatin.}\label{fig:CellLines}
\end{figure}

\subsection{Sensitive/resistant label reversal is common
(Figure~\protect\ref{fig:CellLines})}
The cell line lists and information sources are summarized in
Figure~\ref{fig:CellLines}.
Color changes across rows show that there is at least one
sensitive/resistant label reversal for every drug
checked more than once. The sets of cell lines are different
for most drugs, with the exception of cyclophosphamide/pemetrexed
noted above. The cell lines reported for cyclophosphamide
are a subset of those used for pemetrexed, but running the
cyclophosphamide cell lines through \textttm{binreg} does not
produce the gene list reported. The cell lines producing the
cyclophosphamide gene list are a superset of those used for pemetrexed.
The cisplatin signature is based on 30 cell lines assembled by
\citet{gyorffy06};
the heatmap reported
for temozolomide matches that for cisplatin.


Based on the drug sensitivity information, we believe the
orientations given in the August 2008 lists of cell lines (Potti Corr 2 list)
are correct. Assuming this is the case,
Figure~\ref{fig:CellLines} shows the sensitive/resistant
orientations of the
\citet{salter08}
heatmaps
are correct for T and A but incorrect for F and C.
Heatmap orientations in
\citet{potti06}
are
reversed for T, A and F, and we cannot reproduce
the heatmap for C.
However, sample predictions
shown in both Figure~3A of \citet{potti06}
and
Figure~2B of \citet{salter08}
suggest results better than even for all four drugs
[\citet{potti06}
$p$-values: T${}={}$0.002, F${}={}$0.3, A${}={}$0.024, C${}={}$0.003;
\citet{salter08}
$p$-values: T${}={}$0.07, F${}={}$0.02, A${}={}$0.01, C${}={}$0.02].

\subsection{Summary}
Poor documentation hides the fact that sensitive and resistant
labels are being used inconsistently over time, even though
this direction determines whether the drug should be offered or
withheld.
Full details of the figure assembly are given in Supplementary File~4 [\citet{baggerly09d}].

\section{Discussion}
\subsection{On the nature of common errors}
In all of the case studies examined above, forensic
reconstruction identifies errors
that are
hidden by poor documentation.
Unfortunately, these case studies are illustrative, not
exhaustive;
further problems
similar to the ones detailed above
are described in the
supplementary reports.
The case studies also share other commonalities.
In particular, they illustrate that
\textit{the most common problems are simple}: for example,
confounding in the experimental design (all TET before all FEC),
mixing up the gene labels (off-by-one errors), and
mixing up the group labels (sensitive/resistant);
most of these mixups involve simple switches or offsets.
These mistakes are easy to make, particularly if working
with Excel or if working with 0/1 labels instead of names
(as with \textttm{binreg}).
We have encountered these and like problems before.
As part of the 2002 Competitive Analysis of Microarray
Data (CAMDA) competition,
\citet{stivers03}
identified and corrected a mixup in annotation affecting
roughly a third of the data which was driven by a simple
one-cell deletion from an Excel file coupled with an
inappropriate shifting up of all values in the affected
column only.
\citet{baggerly04a}, \citet{baggerly04b} and \citet{baggerly05}
describe cases of complete confounding leading to optimistic
predictions for proteomic experiments.
\citet{baggerly08}
describe another
array study where there was a mixup in
attaching sample labels to columns of quantifications,
most likely driven by omission of 3 CEL files leading
to an off-by-three error affecting most of the names.
These experiences and others
make us
worry about
the dual of the italicized statement above, that
\textit{the most simple problems may be common.}

\subsection{On the hiding of simplicity}
While simple mistakes often allow for simple fixes,
incomplete documentation and lack of reproducibility
means that this simplicity is often hidden.
Further, it means that identifying the problems often requires
going across several sources of information, papers and journals,
making simple fixes difficult.
Identifying the training data cell lines in Adria\_ALL.txt
used the fact that they were in the same order as in the
\citet{potti06}
heatmaps for correlation matching.
Identifying cell lines for cisplatin used
knowledge of the possibility of an off-by-one error.
Identifying the overlap of the cell lines used for
cyclophosphamide and pemetrexed required identifying
the cell lines and comparing lists across papers.
Identifying the initial
heatmap for temozolomide required familiarity with the
one for cisplatin. This cascade can lead to a certain
archaeology with respect to what has been
established ``as previously shown.'' This can also
render some conclusions unfalsifiable, for example, with
the rejoinder ``they get our results when they use our methods''
provided without details [\citet{potti07};
see Supplementary File~5, \citet{baggerly09e}].

\subsection{What should be done}
So, what can be done? We address this question
in two parts: with respect to the specific findings
illustrated here, and with respect to reproducibility
in general.

\subsubsection{In the case of these results}
In the case of
these results,
we don't think the approach works. We think stronger
evidence (including worked examples of how it works)
need to be provided before this approach is used to
guide patient allocation in clinical trials.
In the case of the clinical trial noted above,
assuming the general approach works means that
sensitive/resistant label reversal for one of the
drugs (pemetrexed) may actually put patients at risk
by giving guidance at odds with the truth, whereas
assuming the general approach doesn't work means that
little will be learned from the trial or even that
it may provide misleading support for the approach
given the inclusion of genes that shouldn't
be there (ERCC1, ERCC4).

A broader question is whether this approach could
work if applied correctly. We don't think so.
We have tried making
predictions from the NCI60 cell lines when we step
through the process without the errors noted above,
and we get results no better than chance.
We communicated with the authors extensively in
the early phases of our investigation (which we
recommend), and shared the fact that we didn't think it worked
before submitting our initial note [\citet{coombes07}].
Empirically, however, we have reached an impasse,
as the progression to clinical trials attests.

\subsubsection{In the case of reproducibility in general}
In the case of reproducibility in general, journals
and funding agencies already require that raw data
(e.g., CEL files) be made available. We see it as
unavoidable that complete scripts will also eventually
be required. General guidelines outlining the types
of questions authors should be prepared to answer in
detail are discussed in the REporting recommendations
for tumor MARKer prognostic studies (REMARK)
guidelines~[\citet{mcshane05}].

\subsection{What we\textup{'}re doing}
Partially in response to the examples discussed here,
we instituted new operating procedures
within our group, mostly simple things having to do with report
structure. Reports in our group are typically
produced by teams of statistical analysts and faculty
members, and issued to our biological collaborators.
We now require most of our reports to be written
using \textit{Sweave}~[\citet{leisch02}],
a literate programming combination
of \LaTeX\ source and R [\citet{rteam08}]
code (SASweave and odfWeave
are also available) so that we can rerun the reports
as needed and get the same results. Some typical
reports are shown in the supplementary material.
Most of these reports are written by the statistical
analysts, and read over (and in some cases rerun) by
the faculty members. All reports include an explicit
call to sessionInfo to list libraries and versions
used in the analysis. The working directory and the
location of the raw data are also explicitly specified
(in some cases leading to raw data being moved from
personal machines to shared drives). We also check for
the most common types of errors, which are frequently
introduced by some severing of data from its associated
annotation [e.g., using 0 or 1 for sensitive or resistant
instead of using names (noted above), supplying one matrix of data
and another of annotation without an explicit joining
feature, calls to order one column of a set].
R's ability to let us use row and column labels which
it maintains through various data transformations helps.
These steps have improved reproducibility markedly.
We have also assumed a fairly regular report structure
in order to enhance clarity, including an executive
summary (at most two pages of text) detailing Background,
Methods, Results and Conclusions in the format most familiar
to our biological collaborators. Discussing the Background
and Methods with the collaborators ahead of time helps
ensure we're working on the problems of actual interest.
Finally, we are trying to shift more frequently to the
use of standardized templates for common analyses
(e.g., two group comparisons for microarray studies).
There is definitely a startup cost in time when shifting
to this approach, but this is often made up when we have
to modify earlier analyses months or years later.

\subsection{The bottom line}
While it may be difficult, we think the examples above show
that this added layer
of documentation is required for high-throughput biology.

\section*{Acknowledgment}
We thank
Aron Eklund for comments about the nature of prodrugs.

\begin{supplement}[id=suppA]
\exhyphenpenalty-10000
\sname{Supplement A}
\stitle{Examining doxorubicin in detail\\}
\slink[doi]{10.1214/09-AOAS291SUPPA}
\slink[url]{http://lib.stat.cmu.edu/aoas/291/supplement-1.zip}
\sdatatype{.zip}
\sdescription{Zipped pdf report describing the identification of
ties and sensitive/resistant status for samples checked for
doxorubicin.}
\end{supplement}

\begin{supplement}[id=suppB]
\exhyphenpenalty-10000
\sname{Supplement B}
\stitle{Cisplatin and pemetrexed}
\slink[doi]{10.1214/09-AOAS291SUPPB}
\slink[url]{http://lib.stat.cmu.edu/aoas/291/supplement-2.zip}
\sdatatype{.zip}
\sdescription{Zip file containing two pdf reports,
matchingCisplatinHeatmap
and matchingPemetrexedHeatmap,
describing the matching of the respective heatmaps
together with sample and gene identification.}
\end{supplement}

\begin{supplement}[id=suppC]
\exhyphenpenalty-10000
\sname{Supplement C}
\stitle{Examining combination therapy\\}
\slink[doi]{10.1214/09-AOAS291SUPPC}
\slink[url]{http://lib.stat.cmu.edu/aoas/291/supplement-3.zip}
\sdatatype{.zip}
\sdescription{Zip file containing seven pdf reports:
getTestingNumbers, getTestingClinical (for identification of
blocks and confounding),
checkOldCombinationRule, checkDrugSensitivity (for the combination
rules), mapGeneLists and predictCytoxanSensitivity (for the ROC
curves), and checkingCellLines (for the drug sensitivity values).}
\end{supplement}

\begin{supplement}[id=suppD]
\sname{Supplement D}
\stitle{Surveying cell lines}
\slink[doi]{10.1214/09-AOAS291SUPPD}
\slink[url]{http://lib.stat.cmu.edu/aoas/291/supplement-4.zip}
\sdatatype{.zip}
\sdescription{Zip file containing two pdf reports:
enumeratingCellLines, describing assembly of the figure, and
getTrainingCellLines from the study of combination therapy,
for identification of the cell lines needed to produce
the gene list for cyclophosphamide.}
\end{supplement}

\begin{supplement}[id=suppE]
\exhyphenpenalty-10000
\sname{Supplement E}
\stitle{Examining docetaxel in detail\\}
\slink[doi]{10.1214/09-AOAS291SUPPE}
\slink[url]{http://lib.stat.cmu.edu/aoas/291/supplement-5.zip}
\sdatatype{.zip}
\sdescription{Zipped pdf report describing the identification of
ties and sensitive/resistant status for samples checked for
docetaxel.}
\end{supplement}

All reports are in \textit{Sweave.} More reports, data and code, are
at\break
\href{http://bioinformatics.mdanderson.org/Supplements/ReproRsch-All}
     {http://bioinformatics.mdanderson.org/Supplements/ReproRsch-All}.
Reports concerning combination therapy are at
\href{http://bioinformatics.mdanderson.org/Supplements/ReproRsch-Breast}{ReproRsch-Breast}.

\printaddresses


\begin{thebibliography}{99}
\bibitem[\protect\citeauthoryear{Augustine et~al.}{2009}]{augustine09}
\textsc{Augustine, C.~K.},
\textsc{Yoo, J.~S.},
\textsc{Potti, A.},
\textsc{Yoshimoto, Y.},
\textsc{Zipfel, P.~A.},
\textsc{Friedman, H.~S.},
\textsc{Nevins, J.~R.},
\textsc{Ali-Osman, F.} and
\textsc{Tyler, D.~S.}
(2009).
Genomic and molecular profiling predicts response to temozolomide in melanoma.
\textit{Clin. Cancer Res.}
\textbf{15}
502--510. Correction p.~3240.

\bibitem[\protect\citeauthoryear{Baggerly, Coombes and Neeley}{2008}]{baggerly08}
\textsc{Baggerly, K.~A.},
\textsc{Coombes, K.~R.} and
\textsc{Neeley, E.~S.}
(2008).
Run batch effects potentially compromise the usefulness of genomic
signatures for ovarian cancer.
\textit{J. Clin. Oncol.}
\textbf{26}
1186--1187.

\bibitem[\protect\citeauthoryear{Baggerly and Coombes}{2009a}]{baggerly09a}
\textsc{Baggerly, K.~A.} and
\textsc{Coombes, K.~R.}
(2009a).
Supplement 1, ``Examining doxorubicin in detail,'' to ``Deriving
chemosensitivity from cell lines: Forensic bioinformatics and reproducible
research in high-throughput biology.''
DOI: \href{http://dx.doi.org/10.1214/09-AOAS291SUPPA}{10.1214/09-AOAS291SUPPA}.

\bibitem[\protect\citeauthoryear{Baggerly and Coombes}{2009b}]{baggerly09b}
\textsc{Baggerly, K.~A.} and
\textsc{Coombes, K.~R.}
(2009b).
Supplement 2, ``Cisplatin and pemetrexed,'' to ``Deriving
chemosensitivity from cell lines: Forensic bioinformatics and reproducible
research in high-throughput biology.''
DOI: \href{http://dx.doi.org/10.1214/09-AOAS291SUPPB}{10.1214/09-AOAS291SUPPB}.

\bibitem[\protect\citeauthoryear{Baggerly and Coombes}{2009c}]{baggerly09c}
\textsc{Baggerly, K.~A.} and
\textsc{Coombes, K.~R.}
(2009c).
Supplement 3, ``Examining combination therapy,'' to ``Deriving
chemosensitivity from cell lines: Forensic bioinformatics and reproducible
research in high-throughput biology.''
DOI: \href{http://dx.doi.org/10.1214/09-AOAS291SUPPC}{10.1214/09-AOAS291SUPPC}.

\bibitem[\protect\citeauthoryear{Baggerly and Coombes}{2009d}]{baggerly09d}
\textsc{Baggerly, K.~A.} and
\textsc{Coombes, K.~R.}
(2009d).
Supplement 4, ``Surveying cell lines,'' to ``Deriving chemosensitivity
from cell lines: Forensic bioinformatics and reproducible research in
high-throughput biology.''
DOI: \href{http://dx.doi.org/10.1214/09-AOAS291SUPPD}{10.1214/09-AOAS291SUPPD}.


\bibitem[\protect\citeauthoryear{Baggerly and Coombes}{2009e}]{baggerly09e}
\textsc{Baggerly, K.~A.} and
\textsc{Coombes, K.~R.}
(2009e).
Supplement 5, ``Examining docetaxel in detail,'' to ``Deriving
chemosensitivity from cell lines: Forensic bioinformatics and reproducible
research in high-throughput biology.''
DOI: \href{http://dx.doi.org/10.1214/09-AOAS291SUPPE}{10.1214/09-AOAS291SUPPE}.

\bibitem[\protect\citeauthoryear{Baggerly et~al.}{2004}]{baggerly04b}
\textsc{Baggerly, K.~A.},
\textsc{Edmonson, S.~R.},
\textsc{Morris, J.~S.} and
\textsc{Coombes, K.~R.}
(2004).
High-resolution serum proteomic patterns for ovarian cancer detection.
\textit{Endocr. Relat. Cancer}
\textbf{11}
583--584.

\bibitem[\protect\citeauthoryear{Baggerly, Morris and Coombes}{2004}]{baggerly04a}
\textsc{Baggerly, K.~A.},
\textsc{Morris, J.~S.} and
\textsc{Coombes, K.~R.}
(2004).
Reproducibility of SELDI-TOF protein patterns in serum: Comparing
datasets from different experiments.
\textit{Bioinformatics}
\textbf{20}
777--785.



\bibitem[\protect\citeauthoryear{Baggerly et~al.}{2005}]{baggerly05}
\textsc{Baggerly, K.~A.},
\textsc{Morris, J.~S.},
\textsc{Edmonson, S.~R.} and
\textsc{Coombes, K.~R.}
(2005).
Signal in noise: Evaluating reported reproducibility of serum proteomic
tests for ovarian cancer.
\textit{J. Natl. Cancer Inst.}
\textbf{97}
307--309.

\bibitem[\protect\citeauthoryear{Bonnefoi et~al.}{2007}]{bonnefoi07}
\textsc{Bonnefoi, H.},
\textsc{Potti, A.},
\textsc{Delorenzi, M.},
\textsc{Mauriac, L.},
\textsc{Campone, M.},
\textsc{Tubiana-Hulin,~M.},
\textsc{Petit, T.},
\textsc{Rouanet, P.},
\textsc{Jassem, J.},
\textsc{Blot, E.},
\textsc{Becette, V.},
\textsc{Farmer, P.},
\textsc{Andre, S.},
\textsc{Acharya, C.~R.},
\textsc{Mukherjee, S.},
\textsc{Cameron, D.},
\textsc{Bergh, J.},
\textsc{Nevins, J.~R.} and
\textsc{Iggo, R.~D.}
(2007).
Validation of gene signatures that predict the response of breast
cancer to neoadjuvant chemotherapy: A substudy of the EORTC 10994/BIG 00-01
clinical trial.
\textit{Lancet Oncol.}
\textbf{8}
1071--1078.

\bibitem[\protect\citeauthoryear{Chang et~al.}{2003}]{chang03}
\textsc{Chang, J.~C.},
\textsc{Wooten, E.~C.},
\textsc{Tsimelzon, A.},
\textsc{Hilsenbeck, S.~G.},
\textsc{Gutierrez, M.~C.},
\textsc{Elledge, R.},
\textsc{Mohsin, S.},
\textsc{Osborne, C.~K.},
\textsc{Chamness, G.~C.},
\textsc{Allred, D.~C.} and
\textsc{O'Connell, P.}
(2003).
Gene expression profiling for the prediction of therapeutic response to
docetaxel in patients with breast cancer.
\textit{Lancet}
\textbf{362}
362--369.

\bibitem[\protect\citeauthoryear{{Clinical Trial NCT00509366}}{2009}]{clintrial09}
\textsc{Clinical Trial NCT00509366}
(2009).
Study using a genomic predictor of platinum resistance to guide therapy
in stage IIIB/IV non-small cell lung cancer (TOP0602).
Available at \url{http://clinicaltrials.gov/ct2/show/NCT00509366}.


\bibitem[\protect\citeauthoryear{Coombes, Wang and Baggerly}{2007}]{coombes07}
\textsc{Coombes, K.~R.},
\textsc{Wang, J.} and
\textsc{Baggerly, K.~A.}
(2007).
Microarrays: Retracing steps.
\textit{Nat. Med.}
\textbf{13}
1276--1277.

\bibitem[\protect\citeauthoryear{Discover}{2007}]{discover07}
\textsc{Discover}
(2007).
The top 6 genetics stories of 2006.
\textit{Discover} \textbf{January}.

\bibitem[\protect\citeauthoryear{Gentleman}{2005}]{gentleman05}
\textsc{Gentleman, R.}
(2005).
Reproducible research: A bioinformatics case study.
\textit{Stat. Appl. Genet. Mol. Biol.}
\textbf{4}
Article 2.
\MR{2138207}

\bibitem[\protect\citeauthoryear{Gentleman and Temple~Lang}{2007}]{gentleman07}
\textsc{Gentleman, R.} and
\textsc{Temple~Lang, D.}
(2007).
Statistical analyses and reproducible research.
\textit{J.~Comput. Graph. Statist.}
\textbf{16}
1--23.
\MR{2345745}

\bibitem[\protect\citeauthoryear{Gy\"{o}rffy et~al.}{2006}]{gyorffy06}
\textsc{Gy\"{o}rffy, B.},
\textsc{Surowiak, P.},
\textsc{Kiesslich, O.},
\textsc{Denkert, C.},
\textsc{Schafer, R.},
\textsc{Dietel, M.} and
\textsc{Lage, H.}
(2006).
Gene expression profiling of 30 cancer cell lines predicts resistance
towards 11 anticancer drugs at clinically achieved concentrations.
\textit{Int. J. Cancer}
\textbf{118}
1699--1712.


\bibitem[\protect\citeauthoryear{Holleman et~al.}{2004}]{holleman04}
\textsc{Holleman, A.},
\textsc{Cheok, M.~H.},
\textsc{den Boer, M.~L.},
\textsc{Yang, W.},
\textsc{Veerman, A.~J.},
\textsc{Kazemier, K.~M.},
\textsc{Pei, D.},
\textsc{Cheng, C.},
\textsc{Pui, C.~H.},
\textsc{Relling, M.~V.},
\textsc{Janka-Schaub, G.~E.},
\textsc{Pieters, R.} and
\textsc{Evans, W.~E.}
(2004).
Gene-expression patterns in drug-resistant acute lymphoblastic leukemia
cells and response to treatment.
\textit{N. Engl. J. Med.}
\textbf{351}
533--542.

\bibitem[\protect\citeauthoryear{Hsu et~al.}{2007}]{hsu07}
\textsc{Hsu, D.~S.},
\textsc{Balakumaran, B.~S.},
\textsc{Acharya, C.~R.},
\textsc{Vlahovic, V.},
\textsc{Walters, K.~S.},
\textsc{Garman, K.},
\textsc{Anders, C.},
\textsc{Riedel, R.~F.},
\textsc{Lancaster, J.},
\textsc{Harpole, D.},
\textsc{Dressman, H.~K.},
\textsc{Nevins, J.~R.},
\textsc{Febbo, P.~G.} and
\textsc{Potti, A.}
(2007).
Pharmacogenomic strategies provide a rational approach to the treatment
of cisplatin-resistant patients with advanced cancer.
\textit{J. Clin. Oncol.}
\textbf{25}
4350--4357.

\bibitem[\protect\citeauthoryear{Ioannidis et~al.}{2009}]{ioannidis09}
\textsc{Ioannidis, J.~P.},
\textsc{Allison, D.~B.},
\textsc{Ball, C.~A.},
\textsc{Coulibaly, I.},
\textsc{Cui, X.},
\textsc{Culhane, A.~C.},
\textsc{Falchi, M.},
\textsc{Furlanello, C.},
\textsc{Game, L.},
\textsc{Jurman, G.},
\textsc{Mangion, J.},
\textsc{Mehta, T.},
\textsc{Nitzberg, M.},
\textsc{Page, G.~P.},
\textsc{Petretto, E.} and \textsc{van Noort, V.}
(2009).
Repeatability of published microarray gene expression analyses.
\textit{Nat. Genet.}
\textbf{41}
149--155.

\bibitem[\protect\citeauthoryear{Leisch}{2002}]{leisch02}
\textsc{Leisch, F.}
(2002).
Dynamic generation of statistical reports using literate data
analysis.
In \textit{Compstat 2002---Proceedings in Computational Statistics}
(W.~H\"{a}rdle and
B.~R\"{o}nz, eds.)
575--580.
Physika Verlag, Heidelberg, Germany.


\bibitem[\protect\citeauthoryear{Li}{2008}]{li08}
\textsc{Li, C.}
(2008).
Automating dChip: Toward reproducible sharing of microarray data
analysis.
\textit{BMC Bioinformatics}
\textbf{9}
231.

\bibitem[\protect\citeauthoryear{Lugthart et~al.}{2005}]{lugthart05}
\textsc{Lugthart, S.},
\textsc{Cheok, M.~H.}, \textsc{den Boer, M.~L.},
\textsc{Yang, W.},
\textsc{Holleman, A.},
\textsc{Cheng, C.},
\textsc{Pui, C.~H.},
\textsc{Relling, M.~V.},
\textsc{Janka-Schaub, G.~E.},
\textsc{Pieters, R.} and
\textsc{Evans, W.~E.}
(2005).
Identification of genes associated with chemotherapy crossresistance
and treatment response in childhood acute lymphoblastic leukemia.
\textit{Cancer Cell}
\textbf{7}
375--386.

\bibitem[\protect\citeauthoryear{McShane et~al.}{2005}]{mcshane05}
\textsc{McShane, L.~M.},
\textsc{Altman, D.~G.},
\textsc{Sauerbrei, W.},
\textsc{Taube, S.~E.},
\textsc{Gion, M.} and
\textsc{Clark, G.~M.}
(2005).
Reporting recommendations for tumor marker prognostic studies
(REMARK).
\textit{J.~Natl. Cancer Inst.}
\textbf{97}
1180--1184.

\bibitem[\protect\citeauthoryear{Potti and Nevins}{2007}]{potti07}
\textsc{Potti, A.} and
\textsc{Nevins, J.}
(2007).
Reply to Microarrays: Retracing steps.
\textit{Nat. Med.}
\textbf{13}
1277--1278.

\bibitem[\protect\citeauthoryear{Potti et~al.}{2006}]{potti06}
\textsc{Potti, A.},
\textsc{Dressman, H.~K.},
\textsc{Bild, A.},
\textsc{Riedel, R.~F.},
\textsc{Chan, G.},
\textsc{Sayer, R.},
\textsc{Cragun, J.},
\textsc{Cottrill, H.},
\textsc{Kelley, M.~J.},
\textsc{Petersen, R.},
\textsc{Harpole, D.},
\textsc{Marks, J.},
\textsc{Berchuck, A.},
\textsc{Ginsburg, G.~S.},
\textsc{Febbo, P.},
\textsc{Lancaster, J.} and
\textsc{Nevins, J.~R.}
(2006).
Genomic signatures to guide the use of chemotherapeutics.
\textit{Nat. Med.}
\textbf{12}
1294--1300.

\bibitem[\protect\citeauthoryear{Potti et~al.}{2008}]{potti08}
\textsc{Potti, A.},
\textsc{Dressman, H.~K.},
\textsc{Bild, A.},
\textsc{Riedel, R.~F.},
\textsc{Chan, G.},
\textsc{Sayer, R.},
\textsc{Cragun, J.},
\textsc{Cottrill, H.},
\textsc{Kelley, M.~J.},
\textsc{Petersen, R.},
\textsc{Harpole, D.},
\textsc{Marks, J.},
\textsc{Berchuck, A.},
\textsc{Ginsburg, G.~S.},
\textsc{Febbo, P.},
\textsc{Lancaster, J.} and
\textsc{Nevins, J.}
(2008).
Corrigendum to ``Genomics signatures to guide the use of
chemotherapeutics.''
\textit{Nat. Med.}
\textbf{14}
889.

\bibitem[\protect\citeauthoryear{Potti, Nevins and Lancaster}{2009}]{uspatent09}
\textsc{Potti, A.}, \textsc{Nevins, J. R.} and \textsc{Lancaster, J. M.}
(2009).
Predicting responsiveness to cancer therapeutics.
U.S. Patent Application 20090105167. Available at
\url{http://www.freepatentsonline.com/y2009/0105167.html}.


\bibitem[\protect\citeauthoryear{R~Development Core Team}{2008}]{rteam08}
\textsc{R~Development Core Team}
(2008).
\textit{R: A Language and Environment for Statistical Computing}.
R Foundation for Statistical Computing, Vienna, Austria.
Available at \url{http://www.R-project.org}.


\bibitem[\protect\citeauthoryear{Riedel et~al.}{2008}]{riedel08}
\textsc{Riedel, R.~F.},
\textsc{Porrello, A.},
\textsc{Pontzer, E.},
\textsc{Chenette, E.~J.},
\textsc{Hsu, D.~S.},
\textsc{Balakumaran,~B.},
\textsc{Potti, A.},
\textsc{Nevins, J.} and
\textsc{Febbo, P.~G.}
(2008).
A genomic approach to identify molecular pathways associated with
chemotherapy resistance.
\textit{Mol. Cancer Ther.}
\textbf{7}
3141--3149.

\bibitem[\protect\citeauthoryear{Ruschhaupt et~al.}{2004}]{ruschhaupt04}
\textsc{Ruschhaupt, M.},
\textsc{Huber, W.},
\textsc{Poustka, A.} and
\textsc{Mansmann, U.}
(2004).
A compendium to ensure computational reproducibility in
high-dimensional classification tasks.
\textit{Stat. Appl. Genet. Mol. Biol.}
\textbf{3}
37.
\MR{2101484}

\bibitem[\protect\citeauthoryear{Salter et~al.}{2008}]{salter08}
\textsc{Salter, K.~H.},
\textsc{Acharya, C.~R.},
\textsc{Walters, K.~S.},
\textsc{Redman, R.},
\textsc{Anguiano, A.},
\textsc{Garman, K.~S.},
\textsc{Anders, C.~K.},
\textsc{Mukherjee, S.},
\textsc{Dressman, H.~K.},
\textsc{Barry, W.~T.},
\textsc{Marcom, K.~P.},
\textsc{Olson, J.},
\textsc{Nevins, J.~R.} and
\textsc{Potti, A.}
(2008).
An integrated approach to the prediction of chemotherapeutic response
in patients with breast cancer.
\textit{PLoS One}
\textbf{3}
e1908.

\bibitem[\protect\citeauthoryear{Stivers et~al.}{2003}]{stivers03}
\textsc{Stivers, D.~N.},
\textsc{Wang, J.},
\textsc{Rosner, G.~L.} and
\textsc{Coombes, K.~R.}
(2003).
Organ-specific differences in gene expression and UniGene annotations
describing source material.
In \textit{Methods of Microarray Data Analysis III}
(K.~Johnson and
S.~Lin, eds.)
59--72.
Kluwer Academic, Boston.




\end{thebibliography}
\end{document}